\documentclass[aps,prl,reprint]{revtex4-2}
\usepackage{amssymb}
\usepackage{amsmath,bm}
\usepackage{graphicx,color}
\usepackage{bbm}
\usepackage[bottom]{footmisc}
\usepackage{multirow}
\usepackage{mathrsfs}

\usepackage{hyperref}
\hypersetup{
    colorlinks = true,
    linkcolor = blue,
    anchorcolor = blue,
    citecolor = magenta,
    filecolor = blue,
    urlcolor = blue
}

\begin{document}
\title{A Novel Method to Probe the Pronounced Growth of Correlation Lengths in an Active Glass-forming Liquids using Elongated Probe}
\author{Anoop Mutneja $^1$, and Smarajit Karmakar $^{1}$}
\affiliation{
$^1$ Tata Institute of Fundamental Research, 36/P, Gopanpally Village, Serilingampally Mandal,Ranga Reddy District,
Hyderabad, Telangana 500107, India
	}

\begin{abstract}
The growth of correlation lengths in equilibrium glass-forming liquids near the glass transition is considered a critical 
finding in the quest to understand the physics of glass formation. These understandings helped us understand various 
dynamical phenomena observed in supercooled liquids. It is known that at least two different length scales exist - one is of 
thermodynamic origin, while the other is dynamical in nature. Recent observations of glassy dynamics in biological and 
synthetic systems where the external or internal driving source controls the dynamics, apart from the usual thermal noise, led 
to the emergence of the field of active matter. A question of whether the physics of glass formation in these active systems 
is also accompanied by growing dynamic and static lengths is indeed timely. In this article, we probe the 
growth of dynamic and static lengths in a model active glass system using rod-like elongated probe particles, 
an experimentally viable method. We show that the dynamic and static lengths in these non-equilibrium systems grow much more rapidly than their passive 
counterparts. We then offer an understanding of the violation of the Stokes-Einstein relation and Stokes-Einstein-Debye 
relation using these lengths via a scaling theory.
\end{abstract}
\maketitle
One of the fascinating properties exhibited by the supercooled liquids is the existence of local patches with heterogeneous 
dynamical properties, known as dynamic heterogeneity (DH) in the literature \cite{Book1}. Many observed properties like 
stretched exponential relaxation, breakdown of Stokes-Einstein (SE) relation, non-Gaussian  Van Hove function, etc., can be 
attributed to the existence of DH\cite{Book1,KDSROPP,PinakiVanHove}. In recent decades, numerous computer simulations 
and experimental studies have linked a growing dynamic length ($\xi_d$)  with DH \cite{KDSROPP,Omega}. There are 
various methods to compute dynamic length in computer simulations. Doing finite size scaling (FSS) of observables 
linked to DH, e.g. the four-point dynamical susceptibility $\chi_4(t)$ \cite{SKPNAS}, Binder cumulant of Van Hove function \cite{PinakiVanHove,BhanuPRE} etc, is one such popular method. Another elegant method is block analysis in which 
scaling analysis is done in a subsystem \cite{Block,BhanuPRE}. Block analysis also has certain advantages over the traditional 
FSS, like the inclusions of number, concentrations, temperature, and other possible fluctuations that are 
absent in the usual FSS, improved statistics, the need for only one large system size, etc. The most important advantage 
is its applicability in colloidal experiments.   At the same time, it is essential to highlight that it is still challenging to 
reach the degree of supercooling attainable in real molecular liquids via computer simulations, even with the recent 
technique of swap Monte-Carlo\cite{Berthier2019}.  
Thus although simulations and colloidal experiments are very
important in the quest to understand the physics of glass transition, appropriate experimental studies on molecular 
glasses are essential.

Another length that comes up while studying various thermodynamic properties of supercooled liquids is often referred 
to as the ``Mosaic scale" ($\xi_s$) in the random first-order transition theory (RFOT) of glass transition \cite{RFOT1,RFOT2}. 
In simulations, static length can be obtained via FSS of the minimum eigenvalue of a configuration's Hessian matrix 
\cite{SKEigen} computed in the energy-minimized inherent state or via the point-to-set  (PTS) method \cite{PTS}, or by FSS of the characteristic relaxation time, $\tau_\alpha$ (see Supplementary Material),
\cite{SKPNAS}. This length, $\xi_s$, grows more slowly than $\xi_d$, suggesting a possible different origin. 
Thus obtaining the length scales in experimentally relevant glass-forming liquids will 
certainly be a major advancement in developing an unified theory of glass formation. 

Measuring growing lengths in experimental systems is extremely difficult. The computer simulations have the liberty to 
track the position of each particle to obtain detailed microscopic information. In contrast, the same cannot be done in actual 
molecular liquids, where the supercooling effects are enormous to emphasize again. There are indirect experimental 
procedures where people have obtained these length scales \cite{BiroliPaper,Berthier2005,Keys}, but they are 
undoubtedly challenging. Recently, there has been a successful effort to get the static length scale by studying the 
change in the dielectric relaxation in supercooled glycerol in the presence of a larger size cosolute particle 
(sorbitol in this case) \cite{BhanuExpt}. In Refs.\cite{AnoopRot,AnoopTrans,Omega}, we proposed a 
novel method to obtain both $\xi_d$ and $\xi_s$ by studying the effect of supercooling on the dynamics of different-sized 
elongated probe particles. The motivation behind doing the same is twofold. The first is its experimental feasibility. 
The vast literature on single-molecule experiments backs that up \cite{ExptOTPSalol1994,WeeksExpt,Blackburn1996,Mackowiak2011,Zondervan2007,Mackowiak2009,Paeng2015}. 
Secondly, doing FSS in experiments is tricky, while the external probes 
of different sizes would give us dynamical information at different coarse-grained lengths, hence serving our purpose. 

The discovery of glass-like behaviour in various biological systems where the internal source of energy and not the thermal 
fluctuations dictates the system's dynamical evolution leads to the emergence of a new field of research known as active matter.
In these systems, the driving forces in constituent elements or particles often come from internal energy dissipation (
self-propulsion) or external driving. Thus they are inherently non-equilibrium in nature. The dynamical behaviour of these systems 
in their stationary states (non-equilibrium steady states) is of significant current interest due to its relevance in various 
systems. The wound healing phenomenon, dynamics inside the cytoplasms \cite{Malinverno2017,Cerbino2021}, flocking and migratory behaviour of animals, birds, 
insects etc., as well as synthetic Janus colloidal assemblies, driven granular fluid \cite{PhysRevLett.113.025701} etc., are a few examples of active matter systems. Many analytical and numerical models  \cite{janssen2019,vijay2007,cugliandolo2019,caprini2020,merrigan2020,chaki2020,activerfot,szamel2016,
saroj2018,activemct,berthier2013} have been studied within the broad framework of statistical mechanics to 
understand the emergent cooperative behaviour of these active matter systems both in dilute and dense limits. 
Recently, Paul et al. \cite{KallolDH} studied such an active glass model via extensive large-scale simulations 
and showed a dramatic five-fold or more increase in the dynamic length scale compared to a similar passive system. This 
massive increase in the DH in the presence of active particles is also experimentally observed \cite{Malinverno2017,Cerbino2021,PhysRevLett.113.025701}. 
In \cite{KallolDH}, all the existing methods to obtain $\xi_d$ are employed to show that a unique dynamic 
length scale controls the dynamics. Thus this system is an interesting model to check whether the same length 
can be probed by looking at the rotational and translational dynamics of a rod-like probe particle with the ambition 
of its subsequent use in ubiquitous out-of-equilibrium experimental situations.

In this work, we studied an active glass model, which has recently been studied extensively \cite{KallolDH,KallolStatic} 
to understand the effect of non-equilibrium active forces on the phenomena of glass transition. We simulated a binary 
glass-forming liquid known as the Kob-Andersen model\cite{KA}. The system contains the larger (A type) and smaller 
(B type) point particles in an 80:20 ratio. We studied a system with $N=50000$ particles, out of which $N_a = 0.1N$ 
of particles are chosen randomly to be active particles. These particles follow run-and-tumble motion with the 
persistence time of $\tau_p=1.0$  in the Lennard-Jones unit. 
The system is studied in the temperature range $T\in[0.340-2.00]$.
In each simulation, $N_r=20$  probe rods are introduced and their position, $r_{cm}$ and orientation, $\hat{u}$ are 
evolved via Newtonian dynamics. 
Further model and simulation details can be found in the Supplementary Materials (SM).
\begin{figure}
\includegraphics[width=.48\textwidth]{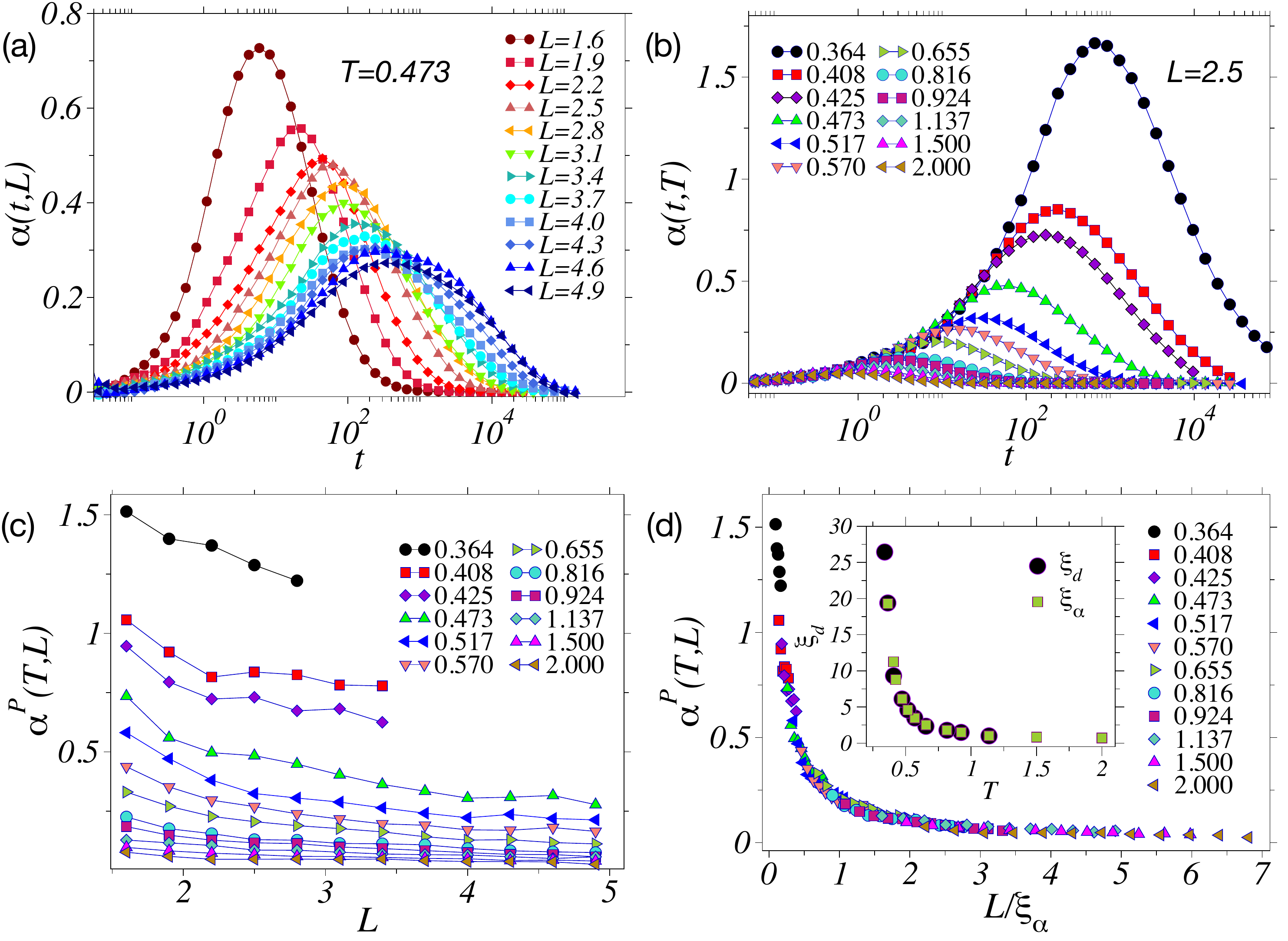}
\caption{(a) Time evolution of the non-normal parameter $\alpha(t,L)$ (Eq:\ref{NNP}) is plotted for rods of variable 
sizes in supercooled system ($T=0.473$).(b) Temperature variation in the time evolution of $\alpha(t,T)$ for rod of length $L=2.5$.
(c) Rod length dependence of the peak height of $\alpha(t)$ for different supercooling temperatures. 
The increase in $\alpha^P(T,L)$ with decreasing temperature and rod length signifies the length scale picked up by the rod dynamics.
(d)The peak heights of panel (c) are collapsed using a scaling relation (see text) to obtain the length scale $\xi_{\alpha}$. The obtained length scale grows by the enormous factor of twenty, and is shown in the inset, along with the comparison with the numbers from Ref.\cite{KallolDH}.}
\label{NNPEvolution}
\end{figure}

Heterogeneity in the dynamics can be quantified via the Van Hove function ($G(r(t^*))$), which is the distribution of 
displacement of particles, $r(t^*)$ during a time interval $t^*$. In supercooled liquids, it is known that the Van Hove function 
has exponential tails on top of the usual Gaussian core owing to DH. The exponential tails correspond to the hopping 
process in the supercooled liquid. A similar correlation function for the rotational degrees of freedom is not Gaussian 
but is the solution to a three-dimensional rotational diffusion equation, as 
\begin{equation}
G(\theta(t^*))=\sum_{n=0}^\infty\frac{2n+1}{2}\mathscr{P}_n(cos(\theta))e^{-n(n+1)D_rt^*}.
\label{diffsoln}
\end{equation}
Here $cos(\theta(t^*))=\hat{u}(t^*)\cdot\hat{u}(0)$, $\mathscr{P}_n$ is $n$th order Legendre polynomials, and $D_r$ is the 
rotational diffusion constant of the rod.  With this solution in hand, one can compute a Non-Normal parameter 
$\alpha(t,T,L)$  as 
\begin{equation}
\alpha(t)=\frac{1}{2}\frac{\left\langle|\hat{u}(t)-\hat{u}(0)|^4\right\rangle}{\left(\left\langle|\hat{u}(t)-\hat{u}(0)|^2\right\rangle\right)^2}+\frac{1}{6}\left\langle|\hat{u}(t)-\hat{u}(0)|^2\right\rangle-1
\label{NNP}
\end{equation}

For a rod in a supercooled liquid, one expects $\alpha(t)$ to peak at the time close to {$\tau_\alpha$} 
and eventually go to zero for large times. Fig.\ref{NNP}(b) shows the time evolution of the non-normal parameter at 
various parent liquid temperatures. Also, for a given supercooling, the effect of DH on the rotational dynamics of the 
rod would dwindle with an increase in rod length because of the increased spatial average. Fig.\ref{NNP}(a) depicts 
the same. The peak height $\alpha^P$ is indeed found to be a good measure for obtaining the dynamic 
heterogeneity length of the parent liquid \cite{AnoopRot} via scaling analysis at equilibrium. Assuming it to be true 
even for these non-equilibrium steady states, we performed a detailed scaling analysis of $\alpha^P$ as a function 
of increasing rod length in our active glass model. Note that $\alpha^P$ should also go to zero in the large rod 
length limit, so there is just one scaling parameter, the length scale $\xi_{\alpha}$. Fig.\ref{NNP}(c) shows the rod length dependence of $\alpha^P(T,L)$ for different temperatures, and on assuming the 
scaling form of $\alpha^P(T,L)=\mathscr{F}(L/\xi_{\alpha}(T))$, one can collapse the data to a master curve. This 
collapse is shown in  Fig.\ref{NNP}(d), along with the comparison of obtained length scale to the dynamic length 
scale of the system obtained from various other methods (from paper \cite{KallolDH}). The remarkable agreement 
of our results  with previously reported one, clearly establishes the robustness of the proposed method in 
out-of-equilibrium active glasses using elongated probe particles. We hope that future experiments will use this 
method to measure these correlations in active glassy systems of experimental relevance.

\begin{figure}[!ht]
\includegraphics[width=0.47\textwidth]{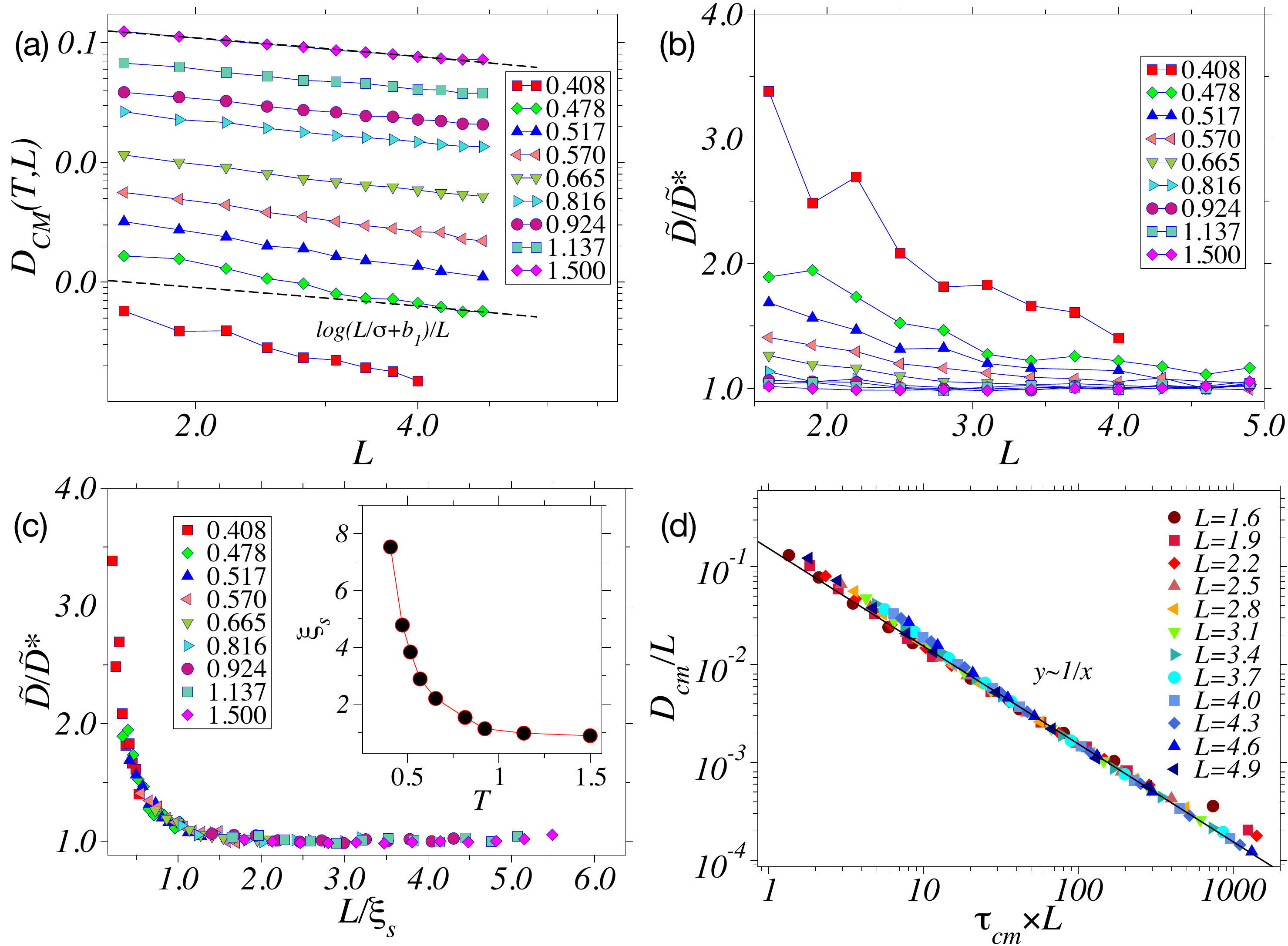}
\caption{(a) The rod length variation of the diffusion constant of CoM of rod in system at different temperatures. The supercooling effects can be seen from the deviation from dotted lines ($D_{cm}\sim ln(L/\sigma+b_1)/L$). (b) To quantify the deviation, $\tilde{D}=D_{cm}L/ln(L/\sigma+b_1)$, scaled by its large rod length limit, $\tilde{D}^*$ is plotted against the rod length. The variation clearly suggest the length scale involved. (c) Data collapse is shown by using the static length scale of the system. (used from ref.\cite{KallolStatic}) (d) $D_{cm}$ and $\tau_{cm}$ follows the inverse relation, thus similar analysis can be performed on $\tau_{cm}$, and is presented in SM.} \label{TauCM}
\end{figure}
\begin{figure*}[!htpb]
\includegraphics[width=0.95\textwidth]{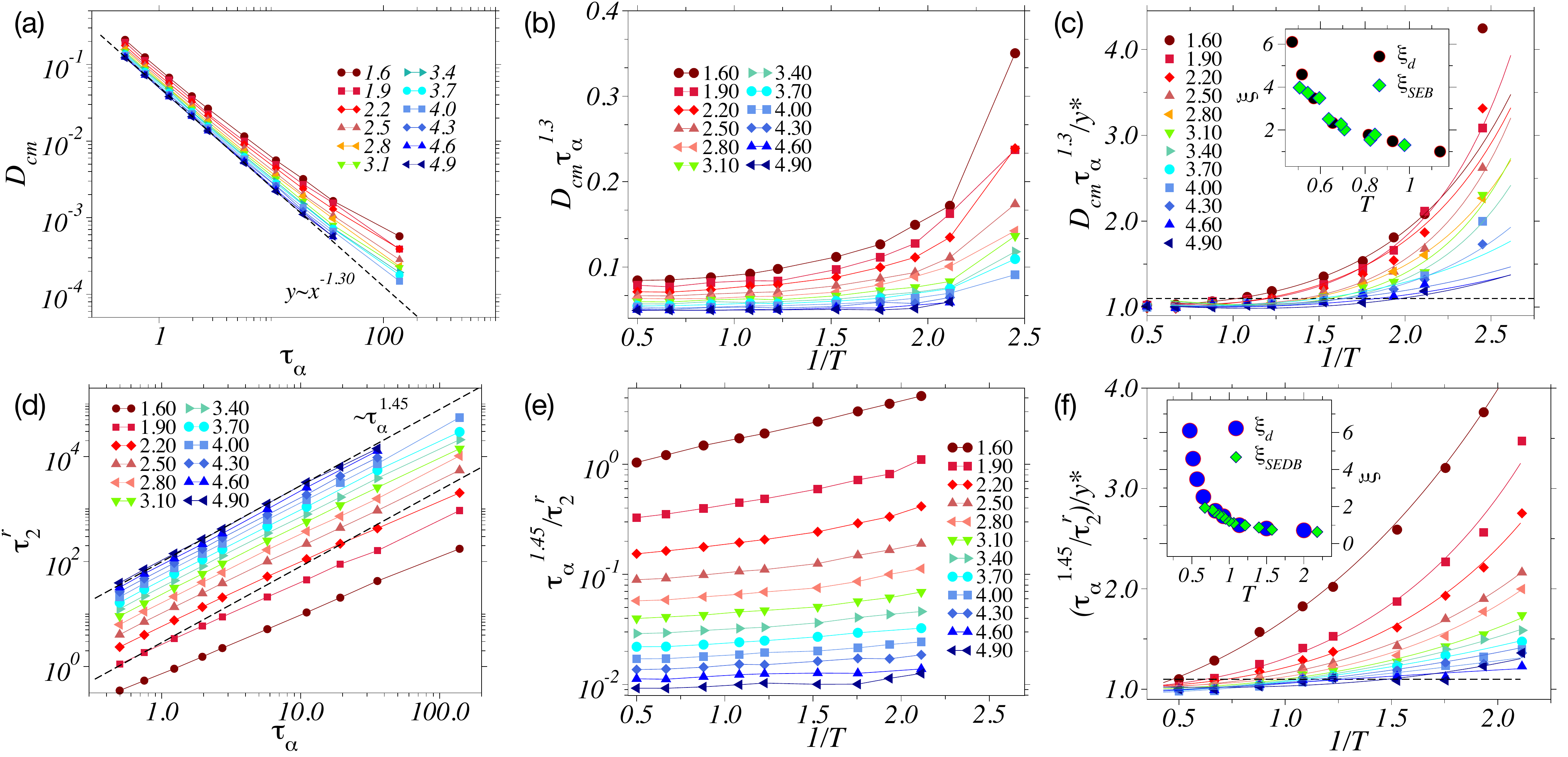}
\caption{(a) Diffusion constant of rod's CoM ($D_{cm}$) shows a power law dependence against $\tau_\alpha$ of the medium with the power of $-1.30$. Shorter rods tend to deviate from this power-law with increasing supercooling (b) The SE parameter, $D_{cm}\tau_\alpha^{1.30}$ is plotted against inverse temperature for different rod sizes. The SEB can be seen to be happening at higher temperatures for smaller rods. (c) The SE parameter scaled by the large $T$ limit is plotted as symbols in main panel. The lines are the fit of these points via function  $y=y^*exp[x+c_0(x/x_0)^n+c_1(x/x_0)^{2n}] $, and SEB temperature is determined as $y/y^*=1.1$ cut. The SEB temperature for different rod lengths is compared with the dynamic length scale of the system in the inset. (d, e, f) Similar analysis shown is for the Stokes-Einstein-Debye (SED) relation using the rotational relaxation time ($\tau_2^r$) of the probe rods. }
\label{SE}
\end{figure*}
After studying the rotational dynamics of the probe particles, we now focus on the translational dynamics of the 
probes. We have calculated the diffusion constant ($D_{cm}$)  and the translational relaxation time ($\tau_{cm}$) of 
the centre of mass (CoM) of the probes. 
It is known that the translational diffusion constant ($D_{cm}$) of a Brownian rod should decrease with increasing 
rod length as $D_{cm}\sim ln(L/\sigma+b_1)/L$, while the relaxation time follows the inverse relation 
$\tau_{cm}\sim L/ln(L/\sigma+b_2)$ \cite{BookPoly}. Here $\sigma=1.0$ is the rod width, and $b_1$, $b_2$ are different constants 
to consider the hydrodynamic interactions. The fitted dotted line in Fig.\ref{TauCM}(a) shows the validation of the same 
for large temperatures ($T=1.5$). One can immediately see the deviation from this behaviour with increasing 
supercooling. The shorter rods tend to move faster than they should have, based on the trend in the large rod limit. 
To quantify the same we studied $\tilde{D}=D_{cm}L/ln(L/\sigma+b_1)$ as shown in Fig.\ref{TauCM}(b). Note that 
the y-axis variation is scaled by large rod length limit $\tilde{D}^*$. On decreasing the temperature, the variables 
$\tilde{D}/\tilde{D}^*$ show stronger deviation from $\tilde{D}/\tilde{D}^*=1$ for smaller rods. This suggests that 
an inherent length scale must be controlling these effects, and one should be able to collapse 
the data to master curves by assuming the scaling relations $\tilde{D}/\tilde{D}^*(L,T)=\mathscr{F}(L/\xi_{D_{cm}}(T))$. 
The collapse obtained in Fig.\ref{TauCM}(c) is indeed very good and the length obtained closely follows the 
temperature variation of the static length scale reported in Ref.\cite{KallolStatic}. This suggests that the method is 
very robust in obtaining static correlation length even in a non-equilibrium system. Moreover, this method has the 
benefit of only requiring the mean translational diffusion constant or the relaxation time of the probes of different 
lengths at different temperatures. Existing results in the literature \cite{ExptOTPSalol1994,WeeksExpt,Blackburn1996,Mackowiak2011,Zondervan2007,Mackowiak2009,Paeng2015} clearly suggests that these measurements 
are not very difficult in real experiments. Since the $D_{cm}$ and $\tau_{cm}$ follows the inverse relation 
(Fig.\ref{TauCM}(d)), similar analysis can be performed on $\tau_{cm}$, and is presented in SM.

Next, we study the Stokes-Einstein (SE) and Stokes-Einstein-Debye (SED) violations in these systems and show how 
the physics of SE and SED breakdown can be understood using the same correlation lengths in a unified manner. SE 
violation is a hallmark of DH. SE relation combines the two dynamic phenomena, namely diffusion and drag, and it 
reads as $D_s\eta/T=const.$. Here $D_s$ is the self-diffusion constant, and $\eta$ is the medium's viscosity. The 
SE relation equates the two relaxation times, the diffusive $R^2/D_s$, and the viscous $\eta R^3/T$, up to a 
constant factor. While in supercooled liquids, it breaks down because different phenomena control different time 
scales. More mobile particles control the diffusive time scale, while the slower set controls the viscous time scale. 
Thus, more DH  in the system would lead to stronger violations in the SE relation. 

The idea of obtaining SEB temperature at various length scales was studied originally in the reference \cite{SastrySE}. 
It was found that if one computes the wave-vector-dependent relaxation time, then there exists a critical wave-vector 
at each temperature below which the SE relation will be obeyed. This critical wave-vector at each temperature turns out 
to be uniquely related to the inverse of the dynamic heterogeneity length scale. A similar crossover is obtained if 
one studies the SE relation for elongated rod-like probe particles with changing probe lengths. This idea was 
successfully used in Ref.\cite{AnoopTrans}  to obtain the largest temperature where SEB happens for a particular rod 
length, and then the corresponding dynamic length could be extracted.

Taking relaxation time is taken as a proxy for viscosity, the SE relation can be rephrased as, 
$D\tau=\frac{f(R,T)}{\pi R}$,
where $R$ would be the effective hydrodynamic radius of the probe rod and $f(R,T)$ would act as a Stokes-Einstein 
breakdown (SEB) parameter. From Fig.\ref{TauCM}(d), we know that the relaxation time and diffusion constant related 
to CoM obey the SE relation, but if one considers the diffusion of a rod with respect to the medium's relaxation times, 
then SE is found to be violated. Fig.\ref{SE}(a) shows the plot of $D_{cm}$ versus $\tau_\alpha$. The larger rods obey 
a modified SE relation as $D_{cm}\tau_{cm}^{1.3}=const.$, where the exponent $1.3$ is not apriori clear to us. Now, if 
we plot the SEB parameter ($D_{cm}\tau_{cm}^{1.3}$) for rods of different sizes (Fig.\ref{SE}(b)), then we will be able 
to obtain the temperature where the breakdown happens. To get the same, we scaled the parameter, 
$D_{cm}\tau_{cm}^{1.3}(T,L)$ by $y^*=D_{cm}\tau_{cm}^{1.3}(T=1.5,L)$ and plotted it against the inverse temperature 
as shown in the main panel of Fig.\ref{SE}(c). One can see the violation of SE relation happening at lower and lower 
temperatures with increasing rod size. The breakdown temperature corresponding to  $D_{cm}
\tau_{cm}^{1.3}(T,L)/y*=1.1$ for each rod length is taken to be the SEB temperature. The symbols in Fig.\ref{SE}(c) 
were fitted to a function $y=y^*exp[x+c_0(x/x_0)^n+c_1(x/x_0)^{2n}] $ (solid lines in Fig.\ref{SE}(c)), to correctly 
obtain the SEB temperature. The obtained SEB temperature plotted against rod length is then compared to the 
dynamic length of the parent liquid in the inset of  Fig.\ref{SE}(c). The two lengths match very well, providing another  
way to obtain the dynamic length from the SE violation of a probe particle. Similarly, for the rotational dynamics of the 
rods, one can check the Stokes-Einstein-Debye (SED) relation. Fig.\ref{SE}(d) shows the rotational relaxation 
($\tau_2^r$, see the definition in SM) time against the $\tau_\alpha$. They follow $\tau_2^r~\tau_\alpha^{1.45}$ 
power law relation in high-temperature and large rod length limit. Note again the origin of the exponent, $1.45$ is 
not immediately known. But, $\tau_\alpha^{1.45}/\tau_2^r$ can then be considered as the SED breakdown (SEDB) 
parameter. SEDB is shown in Fig.\ref{SE}(e) against inverse temperature. After scaling it with the $T=1.5$ value, 
one can see the SEDB happening at lower temperatures for larger rods (Fig.\ref{SE}(f) ). The SEDB temperatures are 
then obtained at $\tau_\alpha^{1.45}/\tau_2^r/y^*=1.1$ cut. The obtained length $\xi_{SEDB}$ is compared with 
the dynamic length of the system in the inset. Again the very good comparison with the dynamical heterogeneity 
length immediately tells us that all these phenomena are  uniquely controlled by the same length scale. Moreover 
SEB and SEDB can then be a good measure to obtain the underlying growth of  the dynamic length scale.

Finally, to conclude, we have obtained the dynamic and static correlation lengths in an active model glass-forming 
liquids using elongated probe particles and show how the rotational dynamics of the probe can capture the dramatic 
growth of the dynamic correlation length, thereby opening a new avenue to obtain the same correlation length in 
various experimental systems including biological systems where glassy dynamics are found to be ubiquitous. Similarly, 
the information of the static length scale using our scaling theory will surely help us in the near future to extend the 
physics of glass transition and the role of various correlation lengths in non-equilibrium systems with similar dynamical 
behaviour. In the end, we show how the physics of Stokes-Einstein and Stokes-Einstein-Debye breakdown can be 
understood using these correlation lengths, as well as how violation of SE and SEB can be elegantly used to even 
obtain the underlying correlation lengths via an unified scaling analysis. We hope these results will encourage 
future experimental measurements in these non-equilibrium active matter systems to probe the growth of correlations 
and help unearth the rich physics of glass transition in a wide variety of systems in equilibrium and non-equilibrium 
situations.

\begin{acknowledgements}	
We acknowledge the support of the Department of Atomic Energy, Government of India, under Project Identification No. RTI 4007. 
SK acknowledges support from Core Research Grant CRG/2019/005373 from Science and Engineering Research Board (SERB) and 
Swarna Jayanti Fellowship grants DST/SJF/PSA-01/2018-19 and SB/SFJ/2019-20/05.
\end{acknowledgements}

\bibliography{draft}

\end{document}


\title{A Novel Method to Probe the Pronounced Growth of Correlation Lengths in an Active Glass-forming Liquids using Elongated Probe\\ Supplementary Material}
\author{Anoop Mutneja $^1$, and Smarajit Karmakar $^{1}$}
\affiliation{
$^1$ Tata Institute of Fundamental Research, 36/P, Gopanpally Village, Serilingampally Mandal,Ranga Reddy District,
Hyderabad, Telangana 500107, India
	}

\maketitle 

\section{Model system and simulation details}
We simulated a binary glass-forming liquid known as the Kob-Andersen model \cite{KA}. The system contains the larger (A type) and smaller (B type) point particles in 80:20 ratio. This famous model mimics the $Ni_{80}P_{20}$ glass-forming liquids. The particles interacts via potential, 
\begin{equation}
V_{\alpha\beta}(\textbf{r})=4\epsilon_{\alpha\beta}
\left[\left(\frac{\sigma_{\alpha\beta}}{r}\right)^{12}-\left(\frac{\sigma
_{\alpha\beta}}{r}\right)^6\right]
\end{equation}
Here $\alpha$ and $\beta$ varies in the particle types, A, and B. The interaction strengths and the radii 
are $\epsilon_{AA}=1.0$, $\epsilon_{AB}=1.5$, $\epsilon_{BB}=0.5$; $\sigma_{AA}=1.0$, $\sigma_{BB}=0.88$ and $\sigma_{AB}=0.8$. The potential is truncated and smoothed upto second order derivative at $r=2.5\sigma_{\alpha\beta}$. The units of length, energy, and time are defined in terms of the diameter of the larger particle ($\sigma_{AA}$), the pre-factor of the potential energy function ($\epsilon_{AA}$), and ($\sqrt{m\sigma_{AA}^2\epsilon_{AA}}$), respectively, where $m=1$ is the mass of the particle.  We studied a system with $N=50000$ particles, out of which $N_a = 0.1N$ of particles are chosen randomly to be active particles. These particles follow run-and-tumble motion with the persistence time of $\tau_p=1.0$  in the Lennard-Jones unit. Physically each $N_a$ particle would get an extra force $F_i^A=f_0(k_x^i\hat{x}+k_y^i\hat{y}+k_z^i\hat{z})$, with $k_{x,y,z}$ are randomly chosen from $\pm1$, after every $\tau_p$ time.The system is studied in the temperature range $T\in[0.340-2.00]$, while the temperature is maintained via the Nos\'{e}-Hoover thermostat.

In each system, $N_r=20$  probe rods are introduced and evolved via Newtonian dynamics. The rods consist of $n_b$ number of beads (A-type) separated by a fixed distance of $0.3$. These beads interact with other particles of the system via the same point potential and add up to the total force and torque of the rod. The $n_b$ varies from $3-14$, which would make the rod length in $L=1.6-4.9$.
\section{The relaxation time, $\tau_\alpha$}
Structural relaxation time, $\tau_\alpha$ is defined as the time when the average value of overlap correlation  $\langle Q(t)\rangle$ function drops to $1/e$ i.e. $\langle Q(\tau_\alpha)\rangle =1/e$.  The overlap correlation function at time $t$ is the fraction of particles till left to move out of their cage	 length $a$. It is defined as follows, 
\begin{equation}
Q(t)=\frac{1}{N}\sum_{i=1}^N\theta\left(a-|\textbf{r}_i(t)-\textbf{r}_i(0)|\right)
\end{equation}
where $\theta(x)$ is the usual step function and value of $a=0.3$ is chosen to 
ignore the decorrelation that might happen due to vibrational motion of 
particles in their cages.

\section{Translational diffusion constant, $D_{cm}$}
The mean squared displacement (MSD) of CoM of the rod is defined as, 
\begin{equation}
MSD=\langle|\textbf{r}_{cm}(t)-\textbf{r}_{cm}(0)|^2\rangle=\langle|\Delta\textbf{r}
_{cm}|^2\rangle,
\end{equation}	
where  $\textbf{r}_{cm}(t)$ is the CoM position vector at a time `$t$'.
Translation diffusion constant ($D_{cm}$) for the CoM of the rod can be obtained 
from MSD using an expression $D_{cm}(T,L)=MSD(t\rightarrow\infty,T,L)/6t$. 
\begin{figure}[!h]
\includegraphics[width=.7\textwidth]{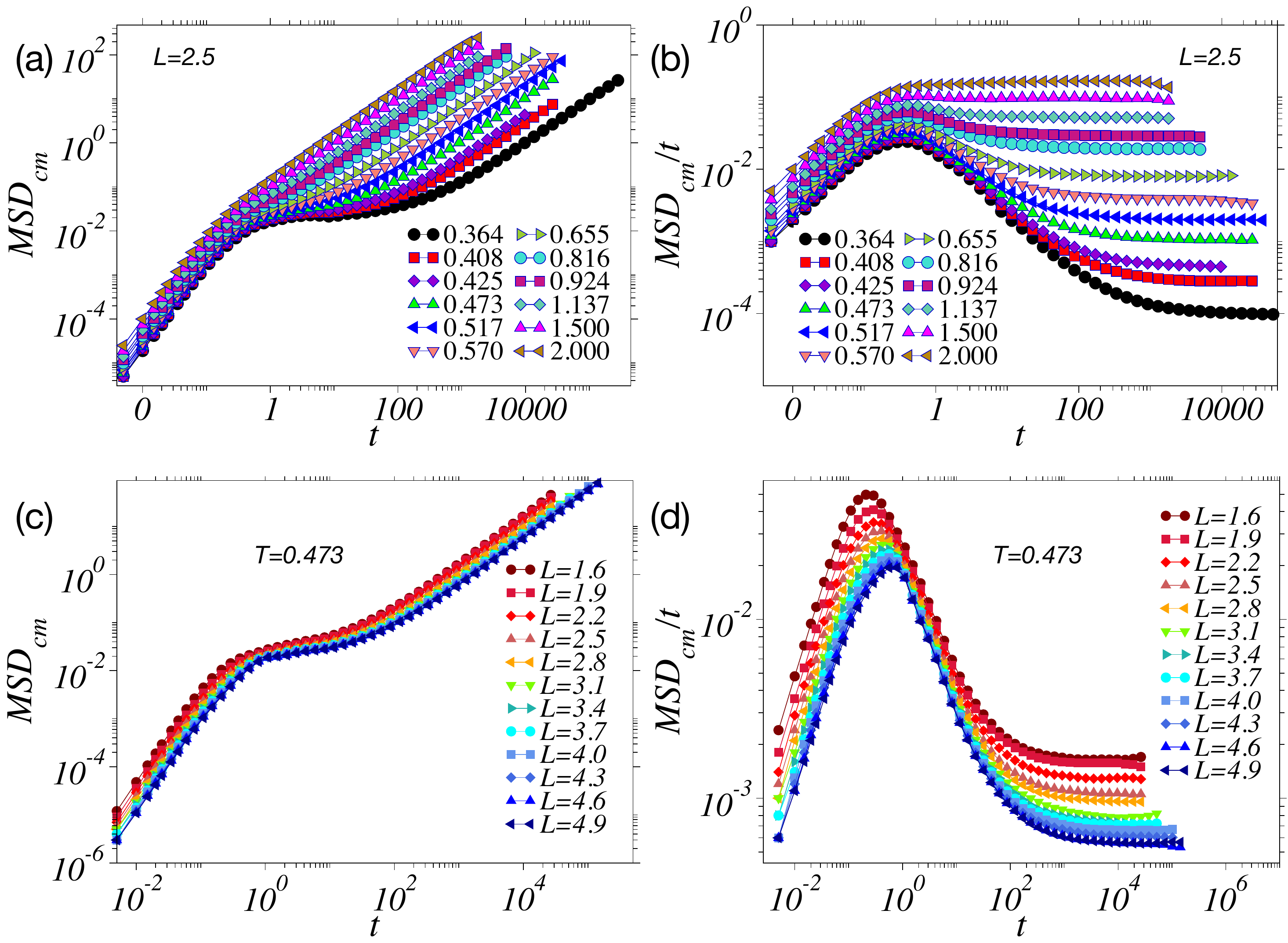}
\caption{(a)Mean squared displacements of CoM of rods of length $L=2.5$ is shown for different parent liquids. Diffusion constant $D_{cm}$ is calculated as the large time limit of $MSD_{cm}/t$. $D_{cm}$ for a rod of length $L=2.5$ immersed in supercooled liquid at different temperatures is shown in panel (b). The rod length dependence of the same can be seen in the lower panels, (c), and (d).    }
\label{MSD}
\end{figure}
\section{Translational relaxation time, $\tau_{cm}$}
The 	translational relaxation time is for CoM of the rod is chosen to be the time when on an average $1/e$ fraction of rods have displaced from their initial positions by a distance of $a=0.3$, just like $\tau_\alpha$.
\begin{figure}[!h]
\includegraphics[width=.7\textwidth]{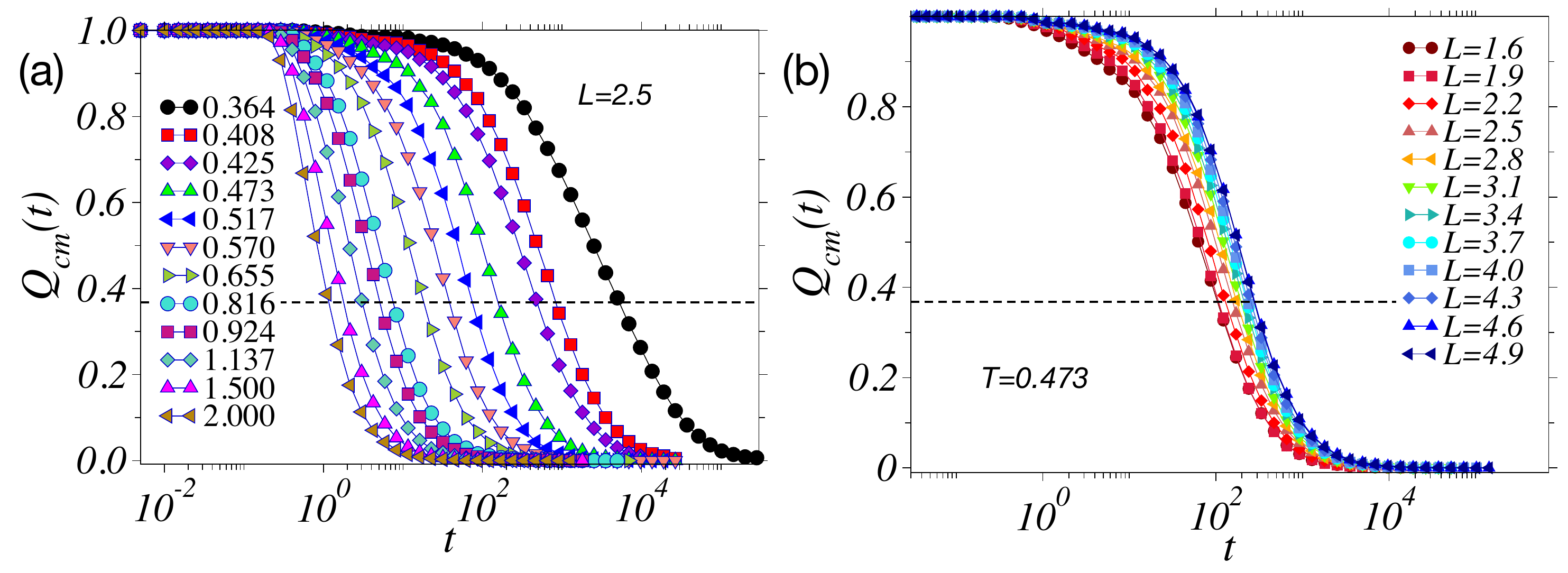}
\caption{(a) The overlap correlation function ($Q_{cm}(t)$) for CoM of the rod of length $L=2.5$ subjected to different supercooling temperatures is shown. The time when averaged $Q_{cm}(t)$ drops to $1/e$ is taken to be the relaxation time ($\tau_{cm}$) of the rod. (b) This panel shows variation in $Q_{cm}(t)$ with changing rod length. }
\label{Qt}
\end{figure}
\section{Rotational relaxation time, $\tau_2^{r}$}
\begin{figure}[!h]
\includegraphics[width=.7\textwidth]{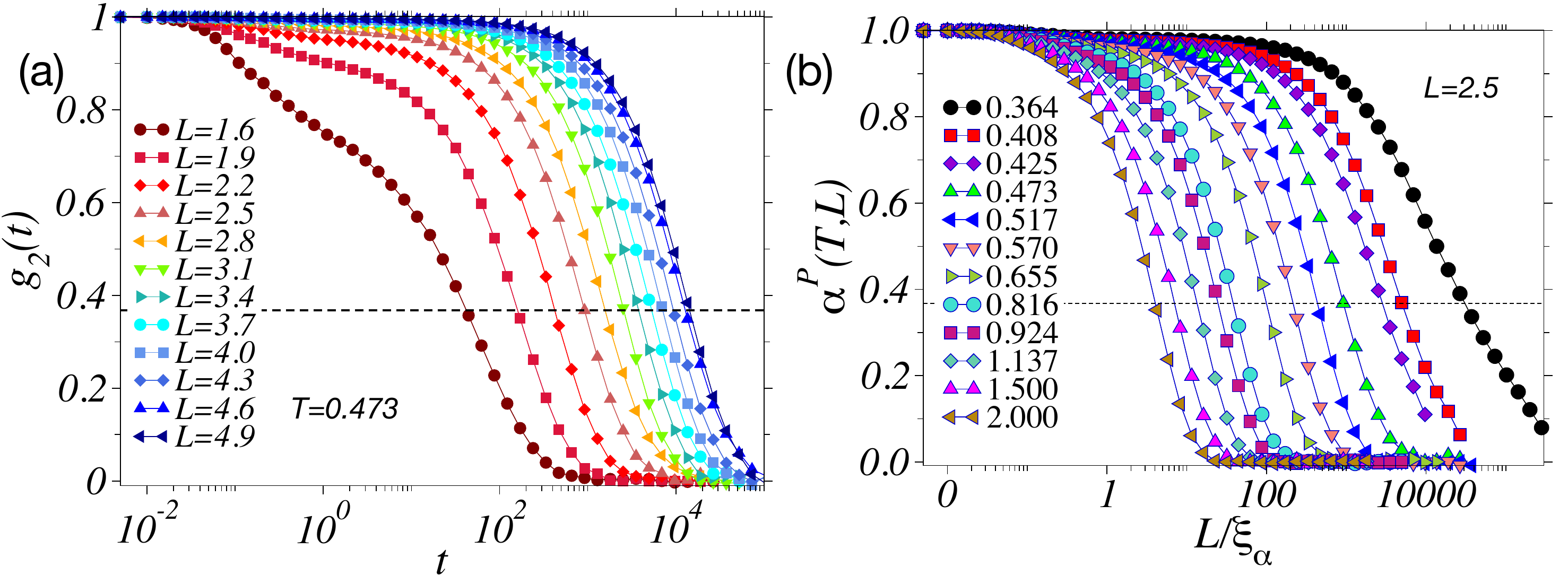}
\caption{(a) Rotational correlation function of rods of different lengths. (b) Rotational correlation function of rod subjected to various supercooling temperatures.}
\end{figure}
The rotational relaxation time is defined as the time when the second order rotational correlation function defined by the following expression decays to $1/e$. 
\begin{equation}
g_2(t)=\langle P_2(\hat{u}(t)\cdot\hat{u}(0))\rangle
\end{equation}
Here, $\hat{u}(t)$ is the orientation vector of a rod at time `$t$'.

\section{Supercooling effects on rod length dependence of translational relaxation time }
\begin{figure}[!h]
\includegraphics[width=.9\textwidth]{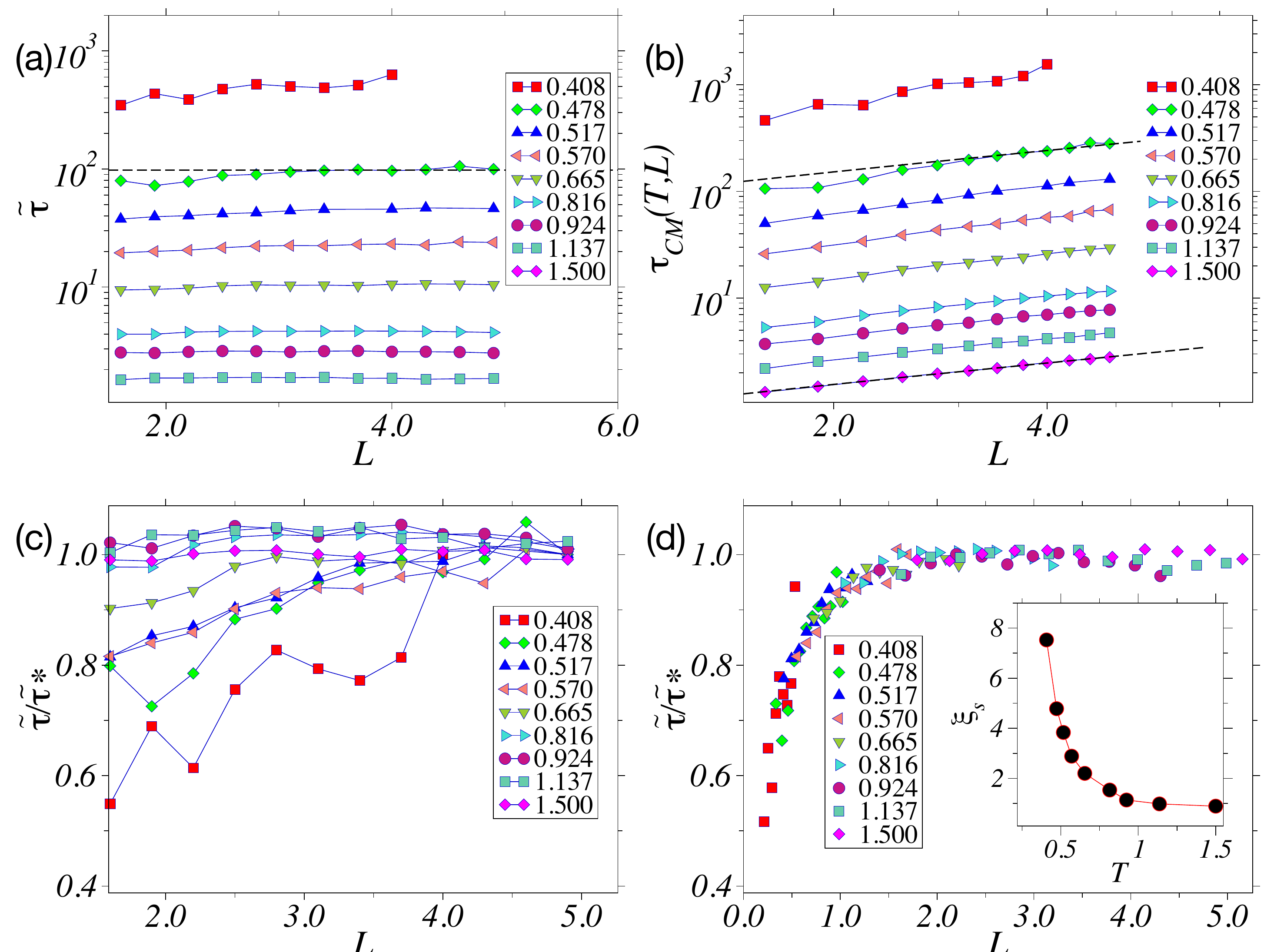}
\caption{(a) The rod length variation of the relaxation time of CoM of rod in system at different temperatures. The supercooling effects can be seen from the deviation from dotted lines ($\tau_{cm}\sim L/ln(L/\sigma+b_2)$). (b) To quantify the deviation, $\tilde{\tau}=\tau_{cm}ln(L/\sigma+b_2)/L$ is presented.(c)$\tilde{\tau}$ scaled by its large rod length limit, $\tilde{\tau}^*$ is plotted against the rod length. The variation clearly suggest the length scale involved. (d) Data collapse is shown by using the static length scale of the system. (used from ref.\cite{KallolStatic})}
\label{TauCM}
\end{figure}
We successfully obtained the static length scale of the active supercooled liquid by studying the deviation of mean CoM diffusion constant from the known behaviour. It is already pointed out in the main text that the CoM diffusion constant and CoM relaxation time for rods are inversely related, implying both contains similar length scale information.  In Fig. \ref{TauCM}(a) we have shown the rod length dependence of $\tau_{cm}$. For large temperature and large rod lengths, they follow the Brownian relation \cite{BookPoly}, $$\tau_{cm}\sim \frac{L}{\ln(L/\sigma+b_2)}$$. Here $\sigma=1.0$ is the rod width, and, $b_2$ is constant  to account for hydrodynamic interactions. It is eminent from Fig. \ref{TauCM}(a) that the shorter rods tend to move faster than expected from large rod length limit. To quantify the same we plotted $\tilde{\tau}=\tau_{cm} ln(L/\sigma+b_2)/L$ in  Fig. \ref{TauCM}(b). In the next step to obtain the supercooling effect we scaled the $\tilde{\tau}$ by its large rod length limit in Fig. \ref{TauCM}(c). The effect of supercooling for various sized rods is clearly different, signalling the involvement of an underlying length scale. The points in \ref{TauCM}(c) are then collapsed to a master curve using the structural length scale reported in Ref. \cite{KallolStatic}.

\bibliographystyle{apsrev4-2}
\bibliography{supli.bib}